\documentclass[epj,referee]{svjour}  
\usepackage{amsmath}
\usepackage{amsfonts,amssymb}
\usepackage{graphicx}

\journalname{European Physical Journal B}

\begin{document}

\title{Matrix Product States for Quantum Many-Fermion Systems}
\titlerunning{Matrix Product States for Quantum Many-Fermion Systems}

\author{Myung-Hoon Chung}
\institute{College of Science and Technology, Hongik University,
Sejong 339-701, Korea \\
mhchung@hongik.ac.kr}
\authorrunning{Myung-Hoon Chung}

\date{Received: date / Revised version: date}
%
\abstract{We describe a simple method to find the ground state
energy without calculating the expectation value of the
Hamiltonian in the time-evolving block decimation algorithm with
tensor network states. For example, we consider quantum
many-fermion systems with matrix product states, which are updated
consistently in a way that accounts for fermion exchange effects.
This method can be applied to a wide class of fermion systems. We
test this method in spinless fermion system where the exact ground
state energy is known. We analyze finite size effects to determine
the ground state energy in the thermodynamic limit that is
compared to the exact value.
}

\maketitle

\section{Introduction}

One of the main challenges in the field of quantum many-fermion
systems is to invent an efficient computational method for finding
the ground states. Up to now, various methods have been proposed
such as exact diagonalization, quantum Monte Carlo, etc. However,
exact diagonalization has limitations in the tractable system
size, while quantum Monte Carlo is plagued by the fermion sign
problem \cite{Troyer}. A practical computational method is the
diffusion Monte Carlo (DMC) \cite{Ceperley,Chung1}, where a set of
replicas is used to represent an approximate ground state. While
the replicas are walking and branching, the number of replicas is
controlled by changing the energy value.

As another accurate computational method without generating random
numbers, the density-matrix renormalization group (DMRG) was
invented by White \cite{White} to simulate strongly correlated
one-dimensional quantum lattice systems. The deeper understanding
of the internal structure of the DMRG is allowed by the matrix
product states (MPS) \cite{Vidal2,Vidal3,Schollwoeck,Orus}. The
method of MPS have attracted much interests for decades in many
different topics \cite{Ostlund,Garcia,Pirvu,Silvi}. Especially,
using MPS, Vidal \cite{Vidal} obtained a simple and fast algorithm
for the simulation of quantum lattice spin systems in
one-dimension. This clever approach is the local updates of
tensors in the MPS by properly handling the Schmidt coefficients.
The concept of local updates is further exploited for quantum
lattice spin systems in two-dimension \cite{Jiang}.

Tensor network states including MPS have been generalized to
describe fermionic systems independently by several groups. The
fermionic projected entangled-pair states
\cite{Kraus,Corboz,Pizorn,Corboz2} was introduced and multiscale
entanglement renormalization ansatz
\cite{Barthel,Corboz3,Corboz4,Pineda,Marti} was generalized to
fermionic lattice systems for the ground states of local
Hamiltonians. These fermionic generalizations share some
similarities, but they also differ in significant ways. Since a
tensor network algorithm is one of variational methods, the
difference between the generalizations can be recognized. It is
remarkable that the infinite projected entangled-pair state for
the ground state in the two-dimensional $t$-$J$ model exhibits
stripes \cite{Corboz5}, which are in contrast to the uniform phase
obtained by other calculations such as variational Monte Carlo and
fixed-node Monte Carlo.

In this paper, getting back to basics for quantum many-fermion
systems, we propose a slightly different algorithm from the
previous approach. We use the concept of updating energy in DMC to
simulate quantum many-fermion systems. We consider the
time-evolving block decimation in imaginary time, including the
energy parameter as in DMC. During the time evolution, we restrict
the accessible states only to MPS. The time evolution of the MPS
is essentially equivalent to the random process of walking and
branching in DMC. The evolution of the MPS norm updates the value
of ground state energy. For spinless fermion systems, we test this
algorithm by comparing the result obtained by our method to the
exact ground state energy. This approach could be a small but
clear step to reach the solution of the two dimensional Hubbard
model, which is one of the current challenging problems. For
future works of tensor network states, we build a user-friendly
library in the scheme of the previous computer code
\cite{Chung2,Chung3}.

\section{Algorithm}

For a quantum system of $N$ sites and $M$ fermions, the typical
dimension of the Hilbert space is determined by the number of
combinations of selecting $M$ elements from $N$ distinct elements.
Thus the dimension of the Hilbert space is an exponential function
of $N$ and $M$. To overcome difficulties arising in the huge
Hilbert space, one may use a small subspace of the Hilbert space:
matrix product states (MPS) or projected entangled-pair states
(PEPS) \cite{Verstraete}. Here we focus on MPS to explain the
algorithm for quantum many-fermion systems.

Representing MPS, we use $N$ three-index tensors
$A^{\sigma_i}_{ab}$ and $N$ Schmidt coefficient vectors
$\lambda_{a}^{i}$, where $i$ runs over all $N$ sites. To describe
the feature of fermions, we impose that $\sigma_i = 0$ or $1$
means vacancy or occupancy at the $i$-th site respectively. For
the bond degree of freedom, the indices $a$ and $b$ run from $0$
to $D-1$. A typical state in the space of matrix product states is
written as
\begin{equation}
|\Psi \rangle = \sum_{\sigma} \sum_{ab\cdots
c}A^{\sigma_0}_{ab}\lambda^{0}_{b} \cdots A^{\sigma_{N-1}}_{ca}
\lambda^{N-1}_{a} | \sigma_{0} \sigma_{1}\cdots \sigma_{N-1}
\rangle ,
\end{equation}
where index-contraction is done as shown in Fig. 1. It is
important to notice one-to-one correspondence between a state of
the spin-like chain represented by $\sigma_{i}$ and a state of the
Fock space written in terms of creation operators
$c^{\dagger}_{i}$ such as
\begin{equation}
| \sigma_{0} \sigma_{1}\cdots \sigma_{N-1} \rangle =
c^{\dagger}_{i_0} c^{\dagger}_{i_1}\cdots c^{\dagger}_{i_{M-1}}|0
\rangle,
\end{equation}
where $\sigma_{i} = 1$ if $i \in \{i_0 , i_1 , \cdots , i_{M-1}
\}$ or $0$ otherwise so that
$\sigma_{0}+\sigma_{1}+\cdots+\sigma_{N-1} = M$.

\begin{figure}
\includegraphics[width= 8cm]{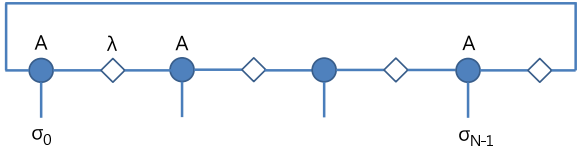}
\caption {Diagrammatic representation of a coefficient in front of
an orthonormal basis in $|\Psi \rangle$. Closed circles and open
diamonds represent tensors $A$ and the Schmidt coefficients
$\lambda$, respectively. For a periodic boundary condition, site
$N-1$ connects back to site $0$. This connection corresponds to
the trace operation. \label{fig:fig1} }
\end{figure}

For a given Hamiltonian $H$, introducing an energy shift $E$ and
the inverse of energy $t$, we consider a formal solution of the
imaginary time Schr\"{o}dinger equation:
\begin{equation}
|\Psi( t ) \rangle = \exp\{ - ( H - E ) t \} |\Psi( 0 )\rangle.
\end{equation}
As $t$ goes to infinity, the state $|\Psi( t ) \rangle$ becomes
the ground state for properly chosen $E$. This is the basic idea
of DMC. The convergence of $E$ is a necessary condition for the
ground state to be stable.

Since we simulate an evolution in the space of MPS, $|\Psi( t )
\rangle$ does not keep a constant number of fermions, while the
Hamiltonian $H$ does not change the number of fermions. To
overcome this problem, we can introduce the chemical potential
$\mu$ and replace $H$ by $H+\mu \sum_i c^{\dagger}_{i} c_{i}$.
However, as far as we are interested in the true ground state, we
do not need $\mu$ because the eventual ground state $|\Psi( \infty
) \rangle$ will sharply peak at some $M$ without $\mu$.

We assume that $H$ is defined by the sum of local operators such
as
\begin{equation}
H = \sum_{\alpha}h_{\alpha}.
\end{equation}
For a given small time step $\tau$, we introduce the
Suzuki-Trotter decomposition as
\begin{equation}
\exp\{ - ( H - E ) t \} \approx \prod^{t/\tau}\prod_{\alpha}
\exp\{(e - h_{\alpha}) \tau \},
\end{equation}
where $e = E / (\sum_{\alpha}1)$ and we should take care of the
order of $\alpha$ for error minimization, for example,
$\prod_{\alpha}=\prod_{\alpha = \text{odd}}\prod_{\alpha =
\text{even}}$. We act the operator $ \prod_{\alpha}\exp\{(e -
h_{\alpha}) \tau \}$ consecutively on the state $|\Psi_n \rangle$
to generate $|\Psi_{n+1} \rangle$, $|\Psi_{n+2} \rangle$, and so
on. The key point is that each output state $\exp\{(e -
h_{\alpha}) \tau \}|\text{MPS}\rangle$, which is outside of the
space of MPS, is approximated into a MPS. As $t$ goes to infinity,
we obtain the approximate ground state in the form of MPS.

\begin{figure}
\includegraphics[width= 8cm]{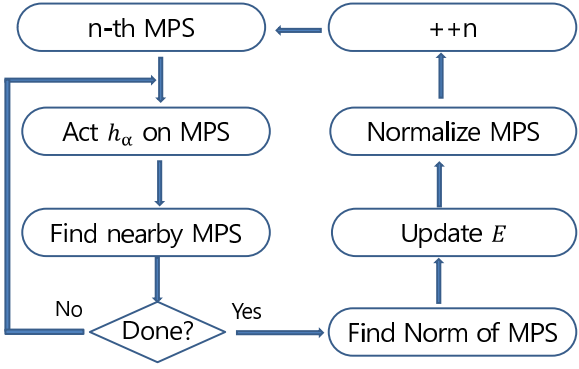}
\caption {Flow chart of the algorithm. \label{fig:fig2} }
\end{figure}

In our algorithm, we start with a normalized MPS $|\Psi_{n}
\rangle$. After we act $ \prod_{\alpha}\exp\{(e - h_{\alpha}) \tau
\}$ on $|\Psi_{n} \rangle$ and find an output MPS
$|\bar{\Psi}_{n+1} \rangle$, we calculate the norm of
$|\bar{\Psi}_{n+1} \rangle$. If the norm is larger (smaller) than
1, we adjust $e$ to become smaller (larger) such as $e_{n+1} =
e_{n} + \xi ( 1 - \langle \bar{\Psi}_{n+1} | \bar{\Psi}_{n+1}
\rangle )$ with a small positive parameter $\xi$. We replace
$|\bar{\Psi}_{n+1} \rangle$ by the normalized one, and call it
$|\Psi_{n+1} \rangle$ for the next iteration. Our algorithm is
summarized in Fig. 2.

\section{One-Body Interaction}

The main step in our algorithm is to find the approximate MPS
after we act $\exp\{(e - h_{\alpha}) \tau \}$ on the previous MPS.
Usually $h_{\alpha}$ is decomposed into a diagonal Hamiltonian
$h^{d}_{\alpha}$ and an off-diagonal Hamiltonian $h^{o}_{\alpha}$
such as
\begin{equation}
\exp\{(e - h_{\alpha}) \tau \} \approx \exp\{(e - h^{d}_{\alpha})
\tau \} \exp( - h^{o}_{\alpha} \tau).
\end{equation}
The diagonal Hamiltonian expands or contracts all basis vectors
without changing direction, while the off-diagonal Hamiltonian
changes both direction and length of bases. For instance, in the
Fock space, the Coulomb repulsion of $V
c^{\dagger}_{i}c_{i}c^{\dagger}_{j}c_{j}$ is a diagonal
Hamiltonian for the basis of occupation number representation. For
an off-diagonal Hamiltonian, there are two cases in physical
interests: one-body interaction,
\begin{equation}
-h^{o}_{\alpha}\tau = d \exp(i\theta)c^{\dagger}_{i}c_{j} + d
\exp(-i\theta)c^{\dagger}_{j}c_{i},
\end{equation}
and two-body interaction,
\begin{equation}
-\tilde{h}^{o}_{\alpha}\tau = d
\exp(i\theta)c^{\dagger}_{i}c^{\dagger}_{j}c_{k}c_{l} + d
\exp(-i\theta)c^{\dagger}_{l}c^{\dagger}_{k}c_{j}c_{i}.
\end{equation}
For simplicity, we here consider the procedure for one-body
interaction only. We will extend our method to two-body
interaction in the future.

With the one-body interaction $-h^{o}_{\alpha}\tau$ of Eq. (7), we
act $\exp(-h^{o}_{\alpha}\tau)$ on a basis vector $ |
\sigma_{0}\cdots \sigma_{i} \cdots \sigma_{j} \cdots \sigma_{N-1}
\rangle$. A simple calculation allows us to find the important
result written in four cases in terms of $\sigma_i = 0$ or $1$ and
$\sigma_j = 0$ or $1$:
\begin{equation}
\left\{
\begin{array}{l}
\exp(-h^{o}_{\alpha}\tau) | \cdots 0 \cdots 0 \cdots \rangle = |
\cdots 0 \cdots 0 \cdots \rangle \nonumber\\
\exp(-h^{o}_{\alpha}\tau) | \cdots 0 \cdots 1 \cdots \rangle =
\cosh d | \cdots 0 \cdots 1 \cdots \rangle \nonumber\\
~~+ \sinh d \exp(i\theta) (-1)^{\sigma_{i+1}+\cdots+\sigma_{j-1}}
| \cdots 1 \cdots 0 \cdots \rangle \nonumber\\
\exp(-h^{o}_{\alpha}\tau) | \cdots 1 \cdots 0 \cdots \rangle =
\cosh d | \cdots 1 \cdots 0 \cdots \rangle \nonumber\\
~~+ \sinh d \exp(-i\theta) (-1)^{\sigma_{i+1}+\cdots+\sigma_{j-1}}
| \cdots 0 \cdots 1 \cdots \rangle \nonumber\\
\exp(-h^{o}_{\alpha}\tau) | \cdots 1 \cdots 1 \cdots \rangle = |
\cdots 1 \cdots 1 \cdots \rangle \nonumber
\end{array} \right .
\end{equation}
It is innovative that the sign of
$(-1)^{\sigma_{i+1}+\cdots+\sigma_{j-1}}$ is based on the fermion
exchange. The above equations are used to update the MPS
represented in new tensors $\tilde{A}$ and new weights
$\tilde{\lambda}$. Since $\exp(-h^{o}_{\alpha}\tau)$ locally
changes $|\Psi \rangle$, the outside tensors and weights remain
unchanged such as $\tilde{A}^{\sigma_k}=A^{\sigma_k}$ for $k < i$
or $k > j$ and $\tilde{\lambda}^{k}=\lambda^{k}$ for $k < i$ or $k
> j - 1$. We should determine $\tilde{A}^{\sigma_l}$ for $i \le l
\le j$ and $\tilde{\lambda}^{l}$ for $i \le l \le j - 1$ by using
the method proposed by Vidal \cite{Vidal}. We first define a
$(2+j-i+1)$-index tensor $M^{\sigma_i \cdots \sigma_j}_{ab}$ such
as
\begin{eqnarray}
M^{\sigma_{i}\sigma_{i+1} \cdots \sigma_{j-1}\sigma_{j}}_{ab}
\equiv \sum_{cd\cdots
fg}\lambda_{a}^{i-1}A^{\sigma_{i}}_{ac}\lambda_{c}^{i}A^{\sigma_{i+1}}_{cd}\lambda_{d}^{i+1}
\nonumber \\
\cdots
\lambda_{f}^{j-2}A^{\sigma_{j-1}}_{fg}\lambda_{g}^{j-1}A^{\sigma_{j}}_{gb}\lambda_{b}^{j}.
\end{eqnarray}
Then we find a single $(2+j-i+1)$-index tensor $\Theta^{\sigma_i
\cdots \sigma_j}_{ab}$ which will be written in the matrix product
form:
\begin{equation}
\left\{
\begin{array}{l}
\Theta^{0\sigma_{i+1} \cdots \sigma_{j-1}0}_{ab} =
M^{0\sigma_{i+1} \cdots \sigma_{j-1}0}_{ab} \nonumber\\
\Theta^{0\sigma_{i+1} \cdots \sigma_{j-1}1}_{ab}= M^{0\sigma_{i+1}
\cdots \sigma_{j-1}1}_{ab}\cosh d \nonumber \\
~~+ M^{1\sigma_{i+1} \cdots \sigma_{j-1}0}_{ab}\sinh d
\exp(-i\theta)
(-1)^{\sigma_{i+1}+\cdots+\sigma_{j-1}} \nonumber\\
\Theta^{1\sigma_{i+1} \cdots \sigma_{j-1}0}_{ab} =
M^{1\sigma_{i+1} \cdots \sigma_{j-1}0}_{ab}\cosh d \nonumber \\
~~+ M^{0\sigma_{i+1} \cdots \sigma_{j-1}1}_{ab}\sinh d
\exp(i\theta)
(-1)^{\sigma_{i+1}+\cdots+\sigma_{j-1}} \nonumber\\
\Theta^{1\sigma_{i+1} \cdots \sigma_{j-1}1}_{ab} =
M^{1\sigma_{i+1} \cdots \sigma_{j-1}1}_{ab} \nonumber
\end{array} \right .
\end{equation}
The new tensors are determined by singular value decomposition
(SVD) of $\Theta$ such as
\begin{equation}
\Theta^{\sigma_i \cdots \sigma_j}_{ab} \approx \sum_{cd\cdots fg}
\bar{A}^{\sigma_i}_{ac}\tilde{\lambda}_{c}^{i}
\tilde{A}^{\sigma_{i+1}}_{cd}\tilde{\lambda}_{d}^{i+1}\cdots
\tilde{A}^{\sigma_{j-1}}_{fg}\tilde{\lambda}_{g}^{j-1}\bar{A}^{\sigma_j}_{gb}.
\end{equation}
To finish updating, we attach the inverse of the Schmidt
coefficients $\tilde{\lambda}$ to the first tensor
$\bar{A}^{\sigma_i}$ and the last tensor $\bar{A}^{\sigma_j}$ in
Eq. (10) such as
\begin{equation}
\tilde{A}^{\sigma_i}_{ac} = \bar{A}^{\sigma_i}_{ac} /
\tilde{\lambda}^{i-1}_{a}, ~~~~ \tilde{A}^{\sigma_j}_{gb} =
\bar{A}^{\sigma_j}_{gb} / \tilde{\lambda}^{j}_{b}.
\end{equation}

Compared to the vigorous calculation in dealing with the
off-diagonal Hamiltonian, the diagonal Hamiltonian is simple and
straightforward to handle. Acting $\exp\{(e - h^{d}_{\alpha})\tau
\}$ on the MPS, we can easily find the relation between old
tensors $A$ and new tensors $\tilde{A}$. In order to update the
MPS, we take the same procedure described in the above: find
$\Theta$, do SVD, and attach $\lambda$.

\section{Spinless Fermion System}

We have tested our method by calculating the ground state energy
and the wave function for spinless fermion system \cite{Rozhkov},
where the exact ground state energy is known. The corresponding
Hamiltonian is written as
\begin{equation}
H = \sum_{i=0}^{N-1}\{-u(c_{i}^{\dagger}c_{i+1} +
c_{i+1}^{\dagger}c_{i}) + v(n_{i} - \frac{1}{2})(n_{i+1} -
\frac{1}{2})\},
\end{equation}
where $n_{i} = c_{i}^{\dagger}c_{i}$ and the periodic boundary
condition is imposed by $c_{N} \equiv c_{0}$.

In this model, the real value $d$ in Eq. (7) is given by the
hopping parameter $u$ times $\tau$, and that the phase $\theta$ in
Eq. (7) is simply equal to zero. When we act
$\exp(-h^{o}_{\alpha}\tau)$ on the MPS, the four-index tensor
$M^{\sigma_{i}\sigma_{i+1}}_{ab}$ in Eq. (9) is involved because
the model has only the nearest neighbor interaction. We follow the
procedure up to Eq. (11) to update the tensors $A^{\sigma_i}_{ab}$
and the weights $\lambda^{i}_{a}$.

Considering the evolution by $\exp\{(e - h^{d}_{\alpha})\tau \}$
in this model with $e=E/N$, we obtain the single four-index tensor
such as
\begin{eqnarray}
\Theta^{\sigma_i \sigma_{i+1}}_{ab}
=\sum_{c}\lambda_{a}^{i-1}A^{\sigma_{i}}_{ac}\lambda_{c}^{i}A^{\sigma_{i+1}}_{cb}
\lambda_{b}^{i+1}\nonumber \\
\times \exp\{[e - v(\sigma_{i} - \frac{1}{2})(\sigma_{i+1} -
\frac{1}{2})]\tau \}.
\end{eqnarray}
Keeping $D$ largest weights $\tilde{\lambda}^{i}_{c}$ among $2D$
values in SVD of $\Theta^{\sigma_i \sigma_{i+1}}_{ab}$, we obtain
the approximate tensor as shown in Eq. (10). We attach the inverse
of the Schmidt coefficients to update the tensors $A$ as shown in
Eq. (11).

\begin{figure*}
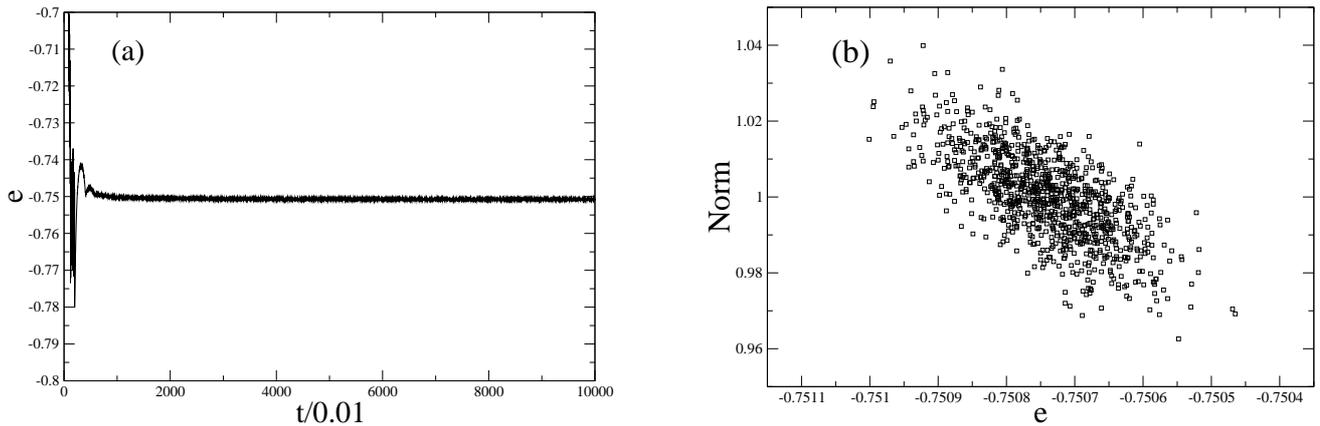

\includegraphics[width= 8cm]{f3a.eps}
\hspace{1cm}
\includegraphics[width= 8cm]{f3b.eps}
\caption {(a) The energy value $e$ as a function of the time
$t/\tau$ for $N=400$ and $D=10$. We let the initial value of $e$
be zero. We find that the MPS is not changed much after $t=10$. It
seems that the MPS is stable forever. (b) The scatter diagram for
the sequence of data $(e_n , \text{Norm}_n )$ from $n=9001$ to
$n=10000$. The thousand points in the scatter diagram shows
negative association as expected. \label{fig:fig3} }
\end{figure*}

\begin{table}
\caption{\label{tab:table1}Numerical results on the average
$\langle e \rangle$, the standard deviation $\delta$ for ground
state energy, and the difference between $\langle e \rangle$ and
the fitted value $\tilde{e}$ in spinless fermion system of $u=1$
and $v=1$.}
\begin{center}
\begin{tabular}{ccccr}
\hline \hline $\tau$ &  $N$  &  $D$  &
$\langle e \rangle \pm \delta$ & $\langle e \rangle - \tilde{e}$\\
\hline
0.02 &  100  &  4         & $-0.75055        \pm 0.00023 $  & $-0.00011$ \\
     &       &  5         & $-0.75190        \pm 0.00025 $  & $ 0.00018$ \\
     &       &  6         & $-0.75272        \pm 0.00021 $  & $ 0.00008$ \\
\hline
0.02 &  200  &  4         & $-0.74856        \pm 0.00019 $  & $ 0.00009$ \\
     &       &  5         & $-0.75039        \pm 0.00022 $  & $-0.00011$ \\
     &       &  6         & $-0.75114        \pm 0.00020 $  & $-0.00013$ \\
\hline
0.01 &  100  &  4         & $-0.75046        \pm 0.00014 $  & $-0.00020$ \\
     &       &  5         & $-0.75175        \pm 0.00012 $  & $ 0.00014$ \\
     &       &  6         & $-0.75262        \pm 0.00011 $  & $-0.00001$ \\
     &       &  10        & $-0.75325        \pm 0.00009 $  & $ 0.00014$ \\
\hline
0.01 &  200  &  4         & $-0.74850        \pm 0.00011 $  & $-0.00003$ \\
     &       &  5         & $-0.75031        \pm 0.00011 $  & $-0.00021$ \\
     &       &  6         & $-0.75108        \pm 0.00010 $  & $-0.00026$ \\
     &       &  10        & $-0.75165        \pm 0.00009 $  & $-0.00005$ \\
\hline
0.01 &  400  &  4         & $-0.74742        \pm 0.00012 $  & $ 0.00015$ \\
     &       &  5         & $-0.74886        \pm 0.00011 $  & $ 0.00035$ \\
     &       &  6         & $-0.74993        \pm 0.00010 $  & $-0.00000$ \\
     &       &  10        & $-0.75074        \pm 0.00008 $  & $-0.00004$ \\
\hline \hline
\end{tabular}
\end{center}
\end{table}

The simulations were performed for the case of $u=1$ and $v=1$ by
fixing $\xi = 0.01$ in the energy update of $e_{n+1} = e_{n} + \xi
( 1 - \text{Norm}_{n+1})$. For given $N$ and $D$ with various
$\tau$, the simulation study shows that the energy $e$ is
converging. Fig. 3(a) shows the convergence of $e$ for
$\tau=0.01$, $N=400$ and $D=10$. We determine the ground state
energy by taking the average $\langle e \rangle$ after annealing
for a long time ($t=90$). Truncation of original states caused by
approximation makes the MPS look like random as shown in Fig.
3(b). The numerical results of the average $\langle e \rangle$ as
the estimates of the ground state energy and the standard
deviation $\delta$ are summarized in Table 1, where we notice
finite size effects. To reduce the finite size effects, we may
need larger values of $N$ and $D$, and a smaller value of $\tau$
as well as a higher-order Suzuki-Trotter decomposition than that
of Eq. (5). Using the results in Table 1, we analyze the finite
size effects with least squares fitting, and determine the fitting
parameters in $\tilde{e}(\tau, N, D)$ for the ground state energy
as
\begin{eqnarray}
\tilde{e}(\tau, N, D)  &=&  -0.74996(7) -0.608(9)\tau^2 \nonumber \\
&-&0.358(3) \frac{1}{N} + 0.214(4) \frac{1}{D^3}.
\end{eqnarray}
The computation time is roughly proportional to $ND^{\alpha}$ with
$\alpha$ ranging from $6$ to $7$.

\section{Conclusion}

Summing up, we have presented an improved time-evolving block
decimation including the energy parameter to obtain the ground
state energy and wave function for quantum many-fermion systems.
If a system has translational symmetry, it is possible to
parallelize local updates \cite{Vidal}. In other words, a single
(or a few) $A$ and $\lambda$ are enough to describe our process.
However, we have not presented this special case here because we
are focusing on a general algorithm, which is applicable to all
cases. We will extend our method to PEPS for two-dimensional
quantum many-fermion systems. The higher-order SVD
\cite{deLathauwer,Levin,Xie} may be useful when we calculate the
norm of PEPS. Furthermore, switching from imaginary time to real
time in Eq. (3) and changing the Hamiltonian slightly, we may
simulate the time evolution from the ground state in order to
explain experimental data of quenching.

\section*{Acknowledgments}
This work was partially supported by Basic Science Research
Program through the National Research Foundation of Korea(NRF)
funded by the Ministry of Education, Science and Technology(Grant
No. 2011-0023395), and by the Supercomputing Center/Korea
Institute of Science and Technology Information with
supercomputing resources including technical support(Grant No.
KSC-2012-C1-09). The author would like to thank K. M. Choi, S. J.
Lee, and J. H. Yeo for helpful discussions.

\end{document}